\documentstyle[12pt,aasms4]{article}
\def\kms{\ifmmode {\rm \ km \ s^{-1}}\else $\rm km \ s^{-1}$\fi}
\def\cm{\ifmmode {\rm \ cm }\else $\rm cm$\fi}
\lefthead{Lee}
\righthead{Polarization Structures in Thomson-Scattered Lines}
\begin{document}
\title{Polarization Structures in the 
Thomson-Scattered Emission Lines in Active Galactic Nuclei}
\author{Hee-Won Lee}
\affil{Dept. of Astronomy and Atmospheric Sciences, 
Kyungpook National University, Taegu, Korea}
\authoremail{hwlee@vega.kyungpook.ac.kr}
\begin{abstract}
A line photon incident in an electron-scattering medium 
is transferred in a diffusive way both in real space and in frequency space,
and the mean number of scatterings changes as the wavelength shifts from
the line center. This leads to the profile broadening and polarization
dependence on the wavelength shift as a function of the Thomson optical 
depth $\tau_T$. We find that the polarization of the Thomson-scattered 
emission lines has a dip around the line center when $\tau_T$ does not exceed 
a few. Various structures such as the polarization flip are also seen. 
An application to an ionized halo component surrounding
the broad emission line region in active galactic nuclei is considered and it
is found that the polarization structures may still persist. 
Brief discussions on observational implications are given.

\end{abstract}

\section{Introduction}

The astrophysical properties of the electron scattering atmosphere 
have been obtained through studies of the polarized radiative transfer 
by many researchers \markcite{cha60, ang69, phi86}(Chandrasekhar 1960, 
Angel 1969, Phillips \& M\'esz\'aros 1986, etc). 
One of the main results from these studies
is that the emergent flux is polarized with a high degree of linear 
polarization up to 11.7 \%
from a very thick plane-parallel atmosphere in the 
direction of the plane. A representative application of this study
has been found in active galactic nuclei (AGN), where the accretion disk
presumed to form the central engine with the supermassive black hole
has a corona component that may be regarded as
a plane-parallel electron scattering atmosphere \markcite{ebs92, rdb90}
(Emmering, Blandford \& Shlosman 1992, Blandford 1990).

Thus far the study of polarization of the Thomson scattered radiation
has been concentrated on the continuum radiation due to the independence
of the Thomson scattering cross section on the frequency. However,
the upper part of the accretion disk is believed to be exposed to the
broad emission line region 
and therefore the Thomson reflected component is expected to 
be present in the broad emission line. 

A line photon incident upon an electron scattering atmosphere will
get a wavelength shift due to the thermal motion of a scatterer.
Assuming that the atmosphere is governed by
a Maxwell-Boltzmann distribution, a typical wavelength shift per scattering
is given by the Doppler width. Therefore, the transfer process is
approximated by a diffusive one both in real space and in frequency space.
It is naturally expected from the random walk nature that the average 
scattering number before escape is smaller in the line center part 
than in the wing part. It is also well known that the polarization of 
the scattered flux is sensitively dependent on the scattering number before 
escape. Therefore, in an electron scattering atmosphere illuminated by
a monochromatic light source, the polarization of the emergent radiation 
will be dependent on the wavelength shift from the line center and may form
a ``polarization structure'', which is expected to be characteristic of 
the Thomson
optical depth of the scattering medium. From this consideration, we may
expect that the polarized flux will have a different profile from 
that of the Thomson-scattered flux.

In this paper, we investigate the polarization of the radiation
both reflected by and transmitted through an electron scattering
slab illuminated by a line-emitting source using a Monte Carlo method
and discuss possible applications to astrophysical sources containing
emission line regions with an electron-scattering atmosphere such as AGN.

\section{ Polarized Radiative Transfer of Thomson-Scattered Lines}

The polarized radiative transfer in an electron-scattering atmosphere
can be simply treated by a Monte Carlo 
method (e.g. \markcite{ang69} Angel 1969). The polarization state associated 
with an ensemble of photons can be described by a density operator 
represented by a $2\times 2$ hermitian density matrix $\rho$. 
\markcite{lbw94} (e.g. Lee, Blandford \& Western 1994).
A Monte Carlo code can be made by recording the density
matrix along with the Doppler shift obtained in each scattering.

In the absence of circular polarization, the density matrix associated with 
the scattered radiation is related
with that of the incident radiation explicitly by
\begin{eqnarray}
\rho'_{11} &=& \cos^2\Delta\phi \  \rho_{11} 
-\cos\theta\sin2\Delta\phi \  \rho_{12} +
\cos^2\theta\sin^2\Delta\phi \  \rho_{22}  \nonumber \\
\rho'_{12} &=&{1\over 2}\cos\theta'\sin2\Delta\phi \  \rho_{11} 
+(\cos\theta\cos\theta'\cos2\Delta\phi +\sin\theta\sin\theta'\cos\Delta\phi) 
\  \rho_{12} \nonumber \\
&& -\cos\theta(\sin\theta\sin\theta'\sin\Delta\phi+
{1\over2}\cos\theta\cos\theta'\sin2\Delta\phi)\  \rho_{22} \\
\rho'_{22} &=& \cos^2\theta' \sin^2\Delta\phi \  \rho_{11} 
+ \cos\theta'(2 \sin\theta\sin\theta'\sin\Delta\phi +
\cos\theta\cos\theta'\sin2\Delta\phi) \  \rho_{12} \nonumber  \\
&& +(\cos\theta\cos\theta'\cos\Delta\phi + \sin\theta\sin\theta')^2
\  \rho_{22} \nonumber
\end{eqnarray}
where the incident radiation is characterized by the wavevector
$\hat{\bf k}_i =(\sin\theta\cos\phi, \sin\theta\sin\phi, \cos\theta)$
and the outgoing wavevector $\hat{\bf k}_f$ is correspondingly given with 
angles $\theta'$ and $\phi'$ with $\Delta\phi = \phi'-\phi$. 
The circular polarization is represented by the imaginary part of the 
off-diagonal element, which is zero in a plane-parallel system and is shown
to be decoupled from the other matrix elements. The
angular distribution of the radiation field is described by the trace
part, from which the scattered wavevector is naturally chosen in the Monte 
Carlo code. 

\section{ Result}

\subsection{Profile formation and polarization of the transmitted component}

The radiative transfer in an electron-scattering atmosphere as a diffusive 
process is studied by \markcite{wey70} Weymann (1970). He considered
a plan-parallel atmosphere with Thomson optical depth $\tau_T$ that embeds
a monochromatic source in the midplane and computed the line profile of 
the emergent flux by adopting the Eddington approximation. 
According to his result, the mean intensity $J$ satisfies the diffusion equation
given by
\begin{equation}
{\partial^2 \over \partial \tau^2} J +{3\over 8}\ {\partial^2\over 
\partial x^2} J = -3\delta(x), 
\end{equation}
where the wavelength shift  $x\equiv (\lambda-\lambda_0)/\Delta\lambda_D$. 
Here, $\Delta\lambda_D =\lambda_0 v_{th}/c$ is the Doppler width and
$\lambda_0$ is the wavelength of the monochromatic source.
 
With the two-stream type boundary conditions, \markcite{wey70}Weymann(1970) 
proposed an approximate solution given by
\begin{equation}
J(x, \tau) = \sum_{n=1}^{\infty} A_n \cos[a_n(\tau-\tau_T/2)] \exp
[-2\sqrt{6}\  a_n |x|/3 ], 
\end{equation}
where $a_n$ is determined from the relation
\begin{equation}
(a_n \tau_T) \tan(a_n \tau_T /2)=\sqrt{3}\ \tau_T ,
\end{equation}
and $A_n$ is obtained from
\begin{equation}
A_n = 4\sqrt{6}\  \tau_T^{-1}\ \sin(a_n\tau_T /2)\  
[(a_n\tau_T)^2 +a_n\tau_T\sin(a_n \tau_T)]^{-1} .
\end{equation}

\placefigure{fig1}
In Fig. 1, by the solid line we plot the profile obtained 
for $\tau_T=8$ investigated by \markcite{wey70} 
Weymann (1970), and by the dotted line we show the corresponding
Monte Carlo results. Here, the radiation source is isotropic located in 
the midplane. The agreement between these two results is good within
1-$\sigma$ except near the center.
The thermal broadening of the profile depends sensitively 
on the Thomson optical depth $\tau_T$. The scattering number contributing
to a given wavelength shift is plotted by the dashed line in the lower panel
and it generally increases monotonically from the line center to the wings.

\placefigure{fig2}
In Fig.~2 we show the profile and the polarization of the reflected
and the penetrated radiation transferred through an
electron-scattering slab illuminated by an anisotropic monochromatic source 
outside the slab, viewed at $\mu=0.5$. 
The horizontal axis represents the wavelength
shift in units of the Doppler width
$\lambda_0 v_{th}/c$ associated with the electronic thermal motion.
Here, the scattering medium is assumed to be governed by a Maxwell-Boltzmann
distribution with temperature $T$ and $\lambda_0$ is the wavelength
of the monochromatic incident radiation.
The Thomson scattering optical depth $\tau_T$ of the slab
is assumed to take values of 0.5, 3, 5.
We first discuss the transmitted radiation.

Because of the random walk nature of the Thomson scattering process, the 
average number of scattering increases as the wavelength shift increases 
toward the wing regime. When $\tau_T\le 1$, near the line center, the emergent
photons are scattered mostly only once. This implies that these
singly scattered photons are mainly contributed from those
incident nearly normally. The photons propagating in the grazing 
direction (with small $\mu_i = \hat{\bf k}_i \cdot \hat{\bf z}$) tend to 
be scattered more than once due to the large Thomson optical 
depth $\tau_T/\mu_i$ in this direction. However, with small $\tau_T$
there is non-negligible contribution from photons with grazing directions.

The resultant polarization is determined from the competition of
the parallel polarization from photons with initially
nearly normal incidence and the perpendicular polarization from
grazingly incident photons. Thus, when $\tau_T\le 1$ a weak 
perpendicular polarization is obtained near the center and as
$\tau_T$ increases the polarization flips to the parallel direction, which
is shown in the case $\tau_T=3$.

On the other hand, the multiply scattered photons are mostly
those with grazing incidence. These photons mainly contribute to the wing 
part. Since the scattering optical depth is small, the scattering plane must 
coincide approximately with the slab plane in a thin atmosphere. 
The polarization develops in the perpendicular
direction to the scattering plane and therefore, the emergent 
photons are polarized in the perpendicular direction to the slab plane,
when the Thomson depth is small.

When $\tau_T\ge 5$, the dependence of the polarization on the
wavelength shift decreases and the overall polarization tends
to lie in the parallel direction to the slab plane. Therefore, the degree
of polarization shows a maximum at the line center, as is shown in the figure. 
The overall parallel polarization is obtained because
the contribution of the singly scattered flux decreases and the
increased mean scattering number before escape leads to an anisotropic
radiation field dominantly in the slab plane direction throughout the
wavelength shifts of the emergent radiation. According to
the Monte Carlo result for a continuum source obtained by \markcite
{ang69} Angel (1969),
when $\tau_T\sim 6$, the polarization reaches the limit of semi-infinite
slab which \markcite{cha60}Chandrasekhar (1960) investigated (see also 
\markcite{la98}Lee \& Ahn 1998).

\subsection{ Polarization of the Reflected Component}

We next discuss the properties of the reflected component.
Firstly, all the reflected components are polarized in the perpendicular
direction with respect to the slab plane for all the scattering optical
depths. Here, one of the most important points to note
is that the linear degree of polarization shows a local minimum
at the line center. The contribution from the singly-scattered photons 
also plays an important role in 
determining the polarization behavior of the reflected component around the 
line center. In the line center, the mean scattering number is also smaller
than in the wing part. Because the light source is assumed to be 
isotropic, the singly-scattered photons constituting the line center
part are contributed almost equally from all the initial directions. Therefore,
the integrated polarization becomes small.

On the other hand, in the wing part, the mean scattering number increases
due to diffusion. The main contribution to the reflected component
is provided by the multiply-scattered photons near the bottom of the slab. 
Therefore, the scattering planes just before reflection are mostly 
coincident with the slab plane, and hence the reflected radiation 
becomes strongly polarized in the perpendicular direction to the slab plane.
As $\tau_T$ increases, the contribution from singly-scattered photons 
decreases and the polarization dip in the center part becomes negligible.

\subsection{Application to an Ionized Galactic Halo}

\markcite{lob98}Loeb (1998) proposed to measure the virial temperature of 
galactic halos using the scattered flux of quasar emission lines. With this 
in mind and for a simple application, we consider a hemispherical halo with the 
Thomson optical depth $\tau_T = 0.1$ with an emission line
source located at the center. It is assumed that the emission source
is anisotropic and illuminate the halo uniformly in the range
$\mu\equiv \cos\theta \ge 0.6$, where $\theta$ is the polar angle.
The observer's line of sight is assumed to have the polar angle
$\theta_l =\cos^{-1} 0.5$, so that the direct flux from the emission source
does not reach the observer. Here, the incident line profile
is chosen to be triangular superimposed with a flat continuum. The half width
of the triangular profile at the bottom is set to be 1 Doppler width
$\Delta\lambda_D$.
This choice is rather arbitrary, and nevertheless considering
the complex profiles and widths that typical quasar emission lines
exhibit, our choice can be regarded as a tolerable approximation.
The line strength is normalized so that the
equivalent width $EW_l = 10 \Delta\lambda_D$.

\placefigure{fig3}
Fig. 3 shows the result, where the linear degree of polarization is
shown by the solid line with 1-$\sigma$ error bars, the scattered flux
by the solid line and the polarized flux with the dotted line. Two local
maxima in the polarization are obtained in the wing parts, and accordingly
the polarized flux possesses larger width than the scattered flux does.
The locations of the polarization maxima are nearly equal to the
Doppler width and therefore, they can be a good measure for the electron
temperature. The slight difference of the widths shown in the
scattered flux and the polarized flux may not be useful to put observational
constraints on the physical properties of the Thomson-scattering medium.
However, the dependence of polarization on the wavelength in
the Thomson-scattered emission lines is quite notable and can provide
complementary information in addition to that possibly obtainable from
the scattered flux profile. 

\section{Observational Implications : AGN Spectropolarimetry}

Spectropolarimetry has been successfully used toward a unified picture
of AGN, according to which, narrow line AGN such as Seyfert 2 galaxies
and narow line radio galaxies are expected to exhibit
the broad emission lines in the polarized flux spectra 
\markcite{am85, ogl97} (Antonucci \& Miller 1985, Ogle et al. 1997).
The ionized component located at high latitude that is responsible for
the polarized broad emission lines is also proposed to give rise to
absorption/reflection features both in X-ray ranges \markcite{kk95}
(Krolik \& Kriss 1995). 

The scattering geometry considered in Fig. 3 may be also applicable to
Seyfert 2 galaxies. However, the typical widths of the broad lines
of order $10000\ {\rm km\ s^{-1}}$ requires very hot scattering
gas of $T\sim 10^7\ {\rm K}$ for a possible polarization
structure considered in this work. The strong
narrow emission lines provide a polarization-diluting component,
which dominates the center part of the broad line features. 
Especially
in the case of hydrogen Balmer lines, atomic effects may also leave
a similar polarization dip in the center part \markcite{ly98}(Lee \& Yun 1998).

Another application of the Thomson scattering process is found
in the upper part of the accretion disk of AGN.
Little or negligible polarization is obtained from a large number of 
polarimetric observations of AGN, which is inconsistent with the
expectation that the emergent radiation from a thick electron-scattering
plane-parallel atmosphere can be highly polarized up to 11.7 percent.
There have been various suggestions including the ideas of corrugated disk 
geometry, magnetic field effects, atomic absorptions \markcite{lnp90, kor98, 
ab96}(e.g. Laor, Netzer \& Piran 1990, Koratkar et al. 1998, Agol \& Blaes 
1996). Negligible polarization is also obtained
when the atmosphere has a small Thomson optical depth $\tau_T \le 3$
\markcite{cht97}(e.g. Chen, Halpern \& Titarchuk 1997). 

It remains
an interesting possibility that the upper part of the disk is illuminated
by the broad emission line sources and therefore the broad lines
may include the Thomson-reflected component from the disk.
The polarization in general will be sensitively dependent on the relative
location of the emission region with respect to the accretion disk.
The origin of the broad emission line region is still controversial
ranging from models invoking a large number of clumpy clouds confined 
magnetically or thermally to an accretion disk wind model \markcite{mc95}
(Murray \& Chiang 1995). The existence and nature
of the outflowing wind around an accretion disk also constitute main 
questions of the unified view of broad absorption line quasars, 
and the polarization of the 
broad lines reflects the importance of resonant scattering 
and electron scattering \markcite{lb97}(e.g. Lee \& Blandford 1997).
Several polarimetric observations reveal some hints of 
polarized broad emission lines \markcite{gm95, coh95}(e.g. Goodrich \&
Miller 1995, Cohen et al. 1995). 

\acknowledgments
This work was supported by the Post-Doc program at Kyungpook National
University. The author is very grateful to the referee Dr. Patrick Ogle
for many helpful suggestions, which improved greatly the presentation
of this paper.

\clearpage
 
\begin{figure}
\caption{Profiles of the emergent radiation from an electron-scattering
plane-parallel atmosphere embedding a monochromatic source at the
mid-plane. The Thomson optical depth is $\tau_T=8$. In the upper panel,
the solid line represents the result which Weymann obtained using a diffusion
approximation and the dotted line shows the Monte Carlo result. The
horizontal axis represents the wavelength shift in units of the
Doppler width $\Delta\lambda_D=\lambda_0 v_{th}/c$.  In the lower panel
the dashed line represents the mean scattering number before escape, which
increases as the wavelength shift increases.
\label{fig1}}
\end{figure}
\begin{figure}
\caption{ Reflected and transmitted radiation in an electron-scattering
plane-parallel atmosphere with Thomson optical depth $\tau_T$ that is
illuminated from a point-like monochromatic source outside the atmosphere.
The line of sight $\hat{\bf k}$ is chosen such that $\mu =
\hat{\bf k}\cdot \hat{\bf z}=\pm 0.5 $, where the normal direction
to the plane is $\hat{\bf z}$, the positive $\mu$ corresponds to
the transmitted component and the negative to the reflected one. 
The polarization direction is denoted by the sign of the degree of
polarization, where a positive degree of polarization implies
polarization in the direction perpendicular to the plane and
a negative in the parallel direction, respectively.
As $\tau_T$ exceeds $\sim 5$, the transmitted component is polarized
in the parallel direction and approaches the result that Chandrasekhar
obtained.
\label{fig2}}
\end{figure}
\begin{figure}
\caption{ The polarization of the Thomson scattered emission lines
from a hemispherical halo, where the emission line source is
located at the center. The halo is illuminated by the line source
uniformly in the range $\mu\equiv\cos\theta \ge 0.6$, where
the polar angle $\theta$ measures the angle from the symmetry axis.
The Thomson optical depth $\tau_T = 0.1$ and the incident line profile
is triangular with half width at the bottom equal to unit Doppler
width superimposed with a flat continuum. The solid line with
1-$\sigma$ error bars represents the linear degree of polarization,
the solid line without error bars represents the scattered flux,
and the dotted line stands for the polarized flux.
\label{fig3}}
\end{figure}

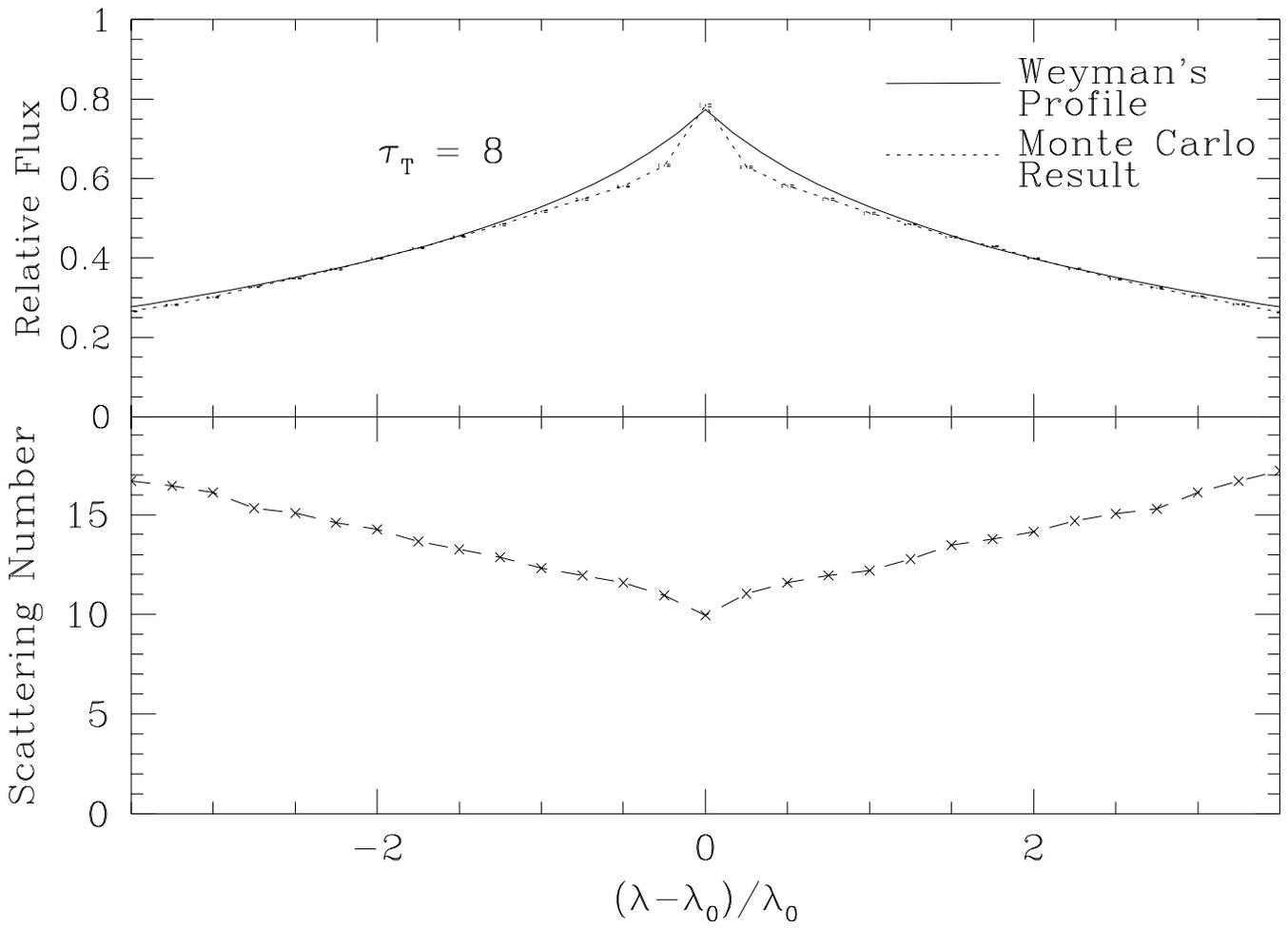
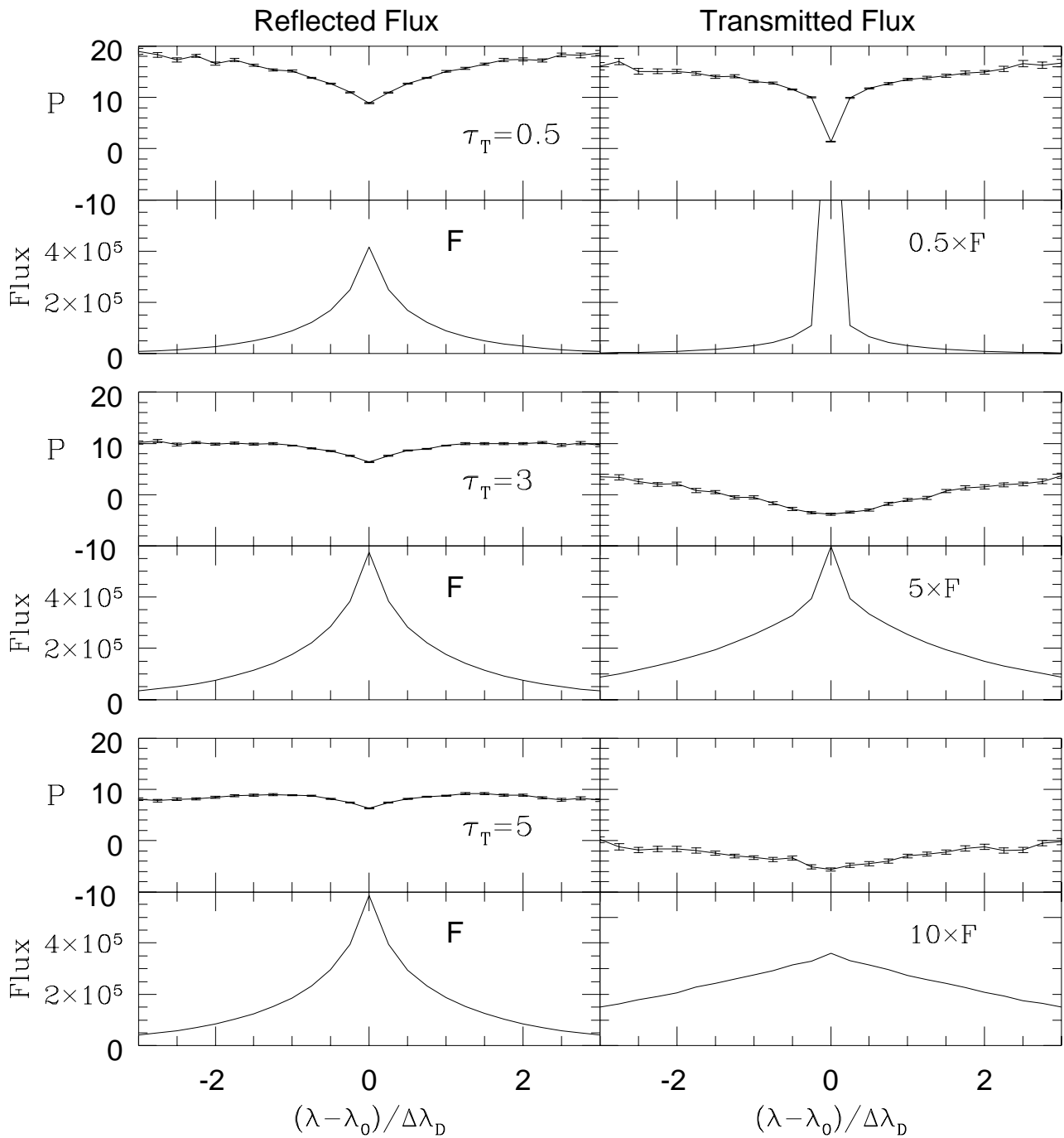
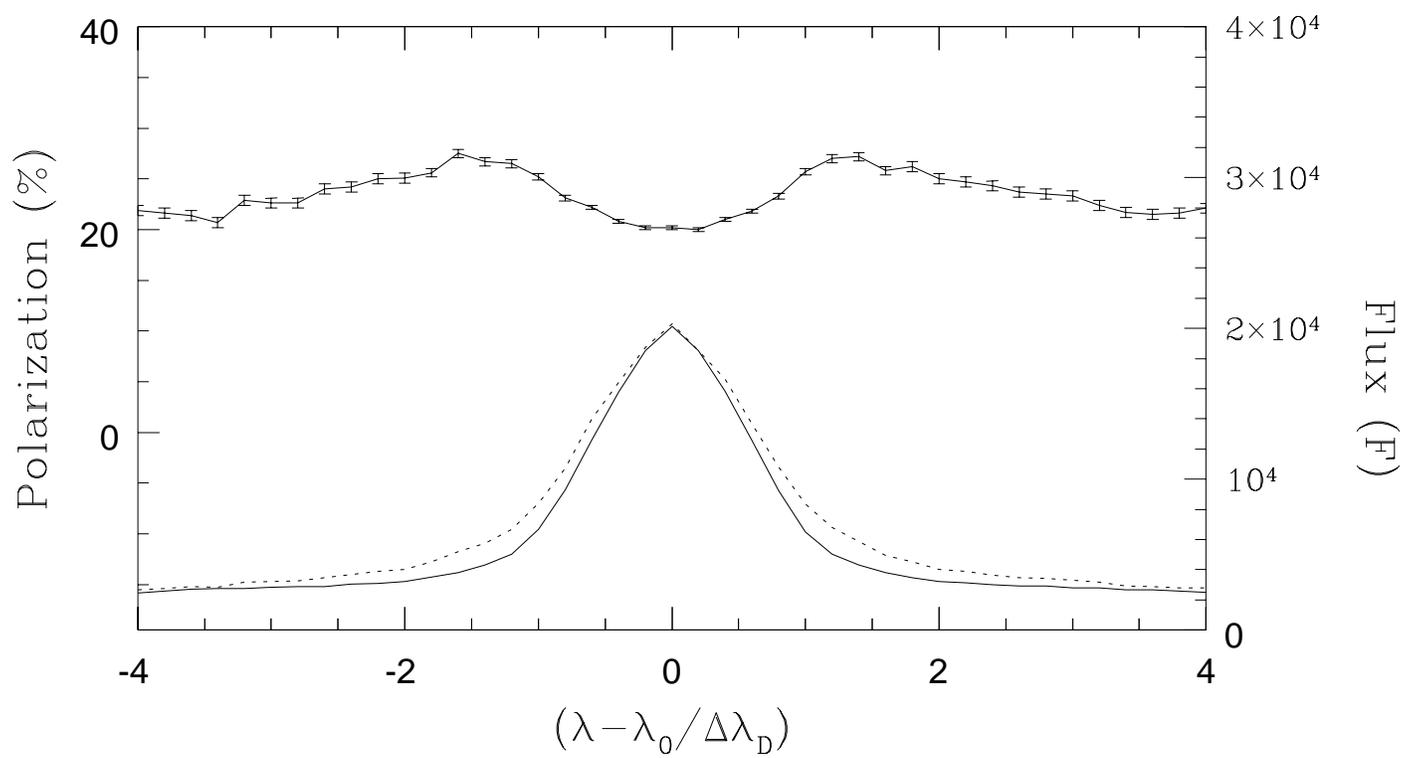
\end{document}